\documentclass[epj]{svjour}
\usepackage{epsfig,amsmath,amssymb,cite}

\newcommand{\kt}{\tilde{\varkappa}}
\newcommand{\ktt}{\tilde{\tilde{\varkappa}}}
\newcommand{\bktt}{\beta \ktt}
\newcommand{\bkt}{\beta \kt}
\newcommand{\ex}{{\rm e}}
\newcommand{\kq}{k^{(q)}}
\newcommand{\ra}{$\rightarrow\;$}

\begin{document}

\title{Phase diagrams of spin-1 Ising model with bilinear
and quadrupolar interactions under magnetic field.
Two-particle cluster approximation}
\titlerunning{Phase diagrams of spin-1 Ising model}

\author{R. R. Levitskii
\and O. R. Baran
\and B. M. Lisnii
}

\institute{
Institute for Condensed Matter Physics,
1 Svientsitskii Str., L'viv-11, 79011, Ukraine
}

\date{\today}
\abstract{
The spin-1 Ising model with bilinear and quadrupolar short-range
interactions under magnetic field is investigated within the
two-particle cluster approximation. It is shown that for those
values of the quadrupolar interaction when at zero magnetic field
the system undergoes a temperature phase transition between
quadrupolar and paramagnetic phases, a triple point may exist in
the temperature vs magnetic field phase diagrams, necessarily
along with a critical point. It is also shown that the critical
points in the temperature vs magnetic field phase diagrams of the
investigated model can be of three types.
}

\maketitle

\section{Introduction}

To obtain an adequate description of several magnetics it does not
suffice to use spin models with bilinear exchange interaction
only, since the exchange interactions of higher-degree spin
operators (the physical origin of which are different for
different magnetic materials) play in them an important role
\cite{Nagaev,Cizmar,Strass}. Hence, a considerable attention is
paid to the effects induced by higher-degree spin terms of an
exchange (e.g. biquadratic) as well as of a non-exchange (e.g.
single-ion anisotropy) origin on the physical properties of the
systems \cite{Chen,Siv2,Siv,On1,Brown,Ru1,Ru2,beg4,Iw2,Iw3}.

One of the simplest model which admits higher-degree spin
exchanges is the spin-1 Ising model ($S_i=S_i^z=0,\pm1$) with
bilinear $K$ and quadrupolar $K^{(q)}$ short-range interactions
\begin{eqnarray}
\label{f1}
H= &-& \sum_{i=1}^N \Gamma S_i \\ \nonumber
&-& \frac{1}{2} \sum_{i,\delta} \left[ K S_i S_{i+\delta} +
K^{(q)} (S_i^2-\frac 23) (S_{i+\delta}^2 -\frac 23) \right] \;\;\;\;
\end{eqnarray}
Here $\Gamma$ is a magnetic field; summation $i,\delta$  is going
over the nearest neighbor pairs.

On the basis of this model in the zero magnetic field case, the
quadrupolar ordering in magnetic materials  was studied
\cite{Chen} within the mean field approximation (MFA). Within this
approximation the quadrupolar moment is the order parameter,
whereas within the constant-coupling approximation or within the
two-particle cluster approximation (TPCA) it is not
\cite{Nagaev,Takahashi1,UFG}. It should be mentioned here, that the
constant-coupling approximation, two-particle cluster
approximation, as well as the Bethe approximation (see e.g. Ref.
\cite{mc3}) yield the same results, which follow,
however, from different considerations.

The model (\ref{f1}) is the partial case of the
Blume-Emery-Griffiths model
which was originally
introduced in Ref. \cite{beg0} to describe the phase
separation and superfluid ordering in He$^3$--He$^4$ mixtures. Now
it is one of the most extensively studied models in the condensed
matter physics. That is so not only because of the relative
simplicity with which approximate calculations for this model can
be carried out and tested, but also because versions and
extensions of the model can be applied for the description of a
wide class of real objects. It proves to be efficient for simple
and multicomponent fluids \cite{beg4,ri1,ri2}, dipolar and
quadrupolar orderings in magnets \cite{Nagaev,Chen,beg4}, binary
alloys  of ferromagnetic and nonmagnetic components
\cite{beg4,alloy1}, ordering in semiconducting alloys
\cite{npd1}. Moreover, due to the richness of its phase diagrams
\cite{mc3,Takahashi2,Hoston,Netz,Balcerzak,Keskin,my3,my4} the
Blume-Emery-Griffiths model is also of a purely theoretical
interest.

In this paper we investigate the magnetic field influence on
thermodynamical characteristics of the model (\ref{f1}) on a
simple cubic lattice at positive values of bilinear and
quadrupolar interactions within the two-particle cluster
approximation. It will be shown that magnetic field not simply
induces a non-zero magnetization in ``paramagnetic'' and
``quadrupolar'' phases, but also may split a temperature phase
transition into a cascade of transitions.

In construction of the phase diagrams of magnetic systems
 with spin-spin and quadrupolar-quadrupolar interactions
\cite{Chen,Siv2,Siv,On1} the consideration is usually restricted
to the  MFA, or it is used as the first-order approximation. For
the Blume-Emery-Griffiths model, the TPCA, in contrast to MFA,
correctly responds to the competition not only between the
ferromagnetic bilinear and negative single-ion anisotropy, but
also between ferromagnetic bilinear and negative biquadratic
interactions. The obtained within TPCA phase diagrams in the
(biquadratic interaction, temperature) plane for the model on
different lattice types in the zero single-ion anisotropy case
\cite{UFG,mc3} qualitatively agree with the Monte-Carlo simulation
results \cite{mc1,mc2}. We would like to mention that the
presented in Refs. \cite{UFG,mc1} phase diagrams in the
temperature vs biquadratic interaction plane at zero single-ion
anisotropy are not complete: the line separating the quadrupole
and staggered quadrupole phases is absent, since only a
one-sublattice model was considered. Furthermore, the fact that
TPCA predictions for the spin-1 Ising model  coincide with the
exact results (see Ref. \cite{Takahashi1}) indicates a high
accuracy of the cluster approximation.

\section{Two-particle cluster approximation}

The free energy of the spin-1 Ising model with bilinear and quadrupolar
short-range interactions under magnetic field (\ref{f1})
within the TPCA reads \cite{UFG,my3,my4}:
\begin{eqnarray}
 \label{f3} && \hspace{-7mm}
F= -k_B T N \left[ (1-z) \ln Z_1 + \frac{z}{2} \ln Z_{12}
   + \frac{2}{9} \beta z K^{(q)} \right] ,
\\ \label{f6} && \hspace{-7mm}
Z_1 = 2 \ex^{ \bkt'} \cdot \cosh( \bkt ) +1 ,
\\ \nonumber && \hspace{-7mm}
Z_{12}= 2 \ex^{ \beta ( 2 \ktt' + K^{(q)} )}
[ \ex^{ \beta K} \cdot \cosh(2 \bktt ) + \ex^{-\beta K} ]
\\ \nonumber &&  \hspace{-7mm} \qquad \quad
+ 4 \ex^{ \bktt' } \cdot \cosh( \bktt ) + 1  . \;\;\;\;
\end{eqnarray}
Here $z$ is the number of nearest neighbors, $\beta=(k_BT)^{-1}$,
$\kt = \Gamma + z \varphi$, $\kt' = -\frac{2}{3}zK^{(q)} + z \varphi'$, $\ktt =
\Gamma + (z-1) \varphi$, $\ktt' = -\frac{2}{3}zK^{(q)} + (z-1) \varphi'$.

For cluster fields $\varphi$ and $\varphi'$
we have the system of equations \cite{UFG,my3,my4}:
\begin{eqnarray}
 && \label{f4}
\frac{ \ex^{ \bkt'} {\cdot} \sinh( \bkt)}{Z_1} =
\\ && \nonumber \quad \!\!\!\!\!
\frac{ \ex^{ \beta (2\ktt'+K^{(q)}+K)} {\cdot} \sinh(2 \bktt)
+ \ex^{ \bktt'} {\cdot} \sinh( \bktt)  }{ Z_{12} }  ,
\\ && \nonumber
\frac{ \ex^{ \bkt'} {\cdot} \cosh( \bkt)}{ Z_1 } =
\\ && \nonumber \quad \!\!\!\!\!
\frac{ \ex^{ \beta (2\ktt'+K^{(q)})} [ \ex^{ \beta K} {\cdot}
\cosh(2 \bktt) + \ex^{- \beta K} ] + \ex^{ \bktt'} {\cdot}
\cosh( \bktt) } { Z_{12} }.
\end{eqnarray}
For magnetization $m=\langle S \rangle$ and quadrupolar moment
$q=3(\langle {S^2} \rangle - \frac{2}{3})$ we have the expressions:
\begin{eqnarray}
\label{f5}
m = \frac{2 \ex^{ \bkt'} {\cdot} \sinh( \bkt)}{Z_1} , \qquad
q = \frac{6 \ex^{ \bkt'} {\cdot} \cosh( \bkt)}{Z_1} - 2.
\end{eqnarray}

\section{Numerical analysis results}

Let us consider results of our numerical investigation of the model (\ref{f1})
on a simple cubic lattice ($z$=6) at positive values of
interactions $K$ and $K^{(q)}$. Here we use the following
notations for the relative quantities: $t\!=3k_BT/(2zK)$,
$h=\Gamma/K$, $\kq\!=K^{(q)}/K$.

It will be easier to describe the effects produces by magnetic
field, if we, at first,  briefly consider and complete the
previous \cite{Nagaev,Chen,Takahashi1,UFG}
results obtained for the case of zero
external magnetic field. Within the TPCA \cite{Nagaev,Takahashi1,UFG},
similarly to the case of MFA \cite{Chen},
we shall distinguish the three following  phases:

\noindent $\bullet$ ferromagnetic phase ($m \ne 0$, $q \ne 0$;
both $m$ and $q$ are convex upwards and decreasing functions of
temperature);

\noindent $\bullet$ paramagnetic phase ($m = 0$, $q \ne 0$,
$q(t\rightarrow\infty)=0$; usually $q(t)$ is a convex downwards
and decreasing or convex upwards and increasing function);

\noindent $\bullet$ quadrupolar phase ($m = 0$, $q \ne 0$;
$q(t)$ is a convex downwards and increasing function).

\noindent Hereinafter, we shall determine what temperature behavior
of the thermodynamic characteristics
is pertinent to a particular phase only at those
values of $\kq$, when we can clearly determine the phase in which
the system is, that is, when a phase transition (PT) takes place
on changing temperature. It should be noted that within the mean
field approximation, the criterion for phase discrimination is
clearer than within the TPCA, since in the MFA not only the
magnetization but also the quadrupolar moment (\ref{f1}) is the
order parameter ($q=0$ in the paramagnetic phase)
\cite{Chen,UFG}.

\begin{figure}
\begin{centering}
{\includegraphics*{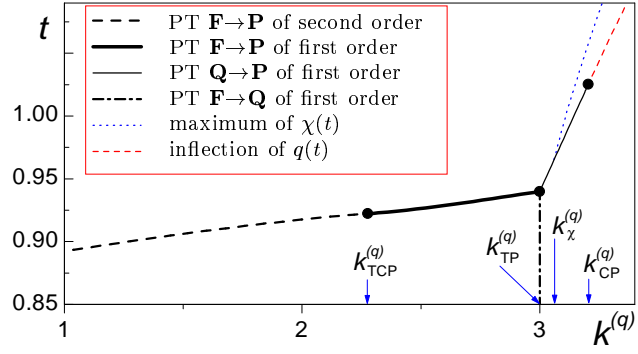}}\\ [1ex]
\end{centering}
\caption{ The $\kq$ vs $t$ phase diagram at zero magnetic field.
Thick dashed and solid lines indicate the PT
ferromagnetic~\ra paramagnetic phase of the second and first
order, respectively. Thin solid line indicates the first order
PT quadrupolar~\ra paramagnetic phase.
Thick dashed-dotted line indicates the first order PT
ferromagnetic~\ra quadrupolar phase. Thin dashed and
dotted lines correspond, respectively, to the inflections in the
temperature dependences of a quadrupolar moment and to the maxima
in the temperature dependences of static magnetic susceptibility.
The special points are the tricritical (TCP), triple (TP)
and critical (CP) points.
$\kq_{\chi}$ is the coordinate of the intersection point
between the temperature phase transition line and the line of the
$\chi(t)$ maxima. } \label{fig1}
\end{figure}

In figure~\ref{fig1} we show the obtained within the TPCA phase
diagram in the $(\kq, t)$ plane. At $\kq < \kq_{\rm TCP}$ (TCP is
the tricritical point ; $\kq_{\rm TCP} = 2.28$) the TPCA yields
the second order temperature PT ferromagnetic~\ra
paramagnetic phase. At $\kq_{\rm TCP} < \kq < \kq_{\rm TP}$
(TP is the triple point; $\kq_{\rm TP}=3.0$) the first order
temperature phase transitions ferromagnetic~\ra paramagnetic phase
take place. At $\kq_{\rm TP} < \kq < \kq_{\rm CP}$ (CP is the
critical point; $\kq_{\rm CP}=3.2$) the first order temperature
PT quadrupolar~\ra paramagnetic phase take place. At $\kq
> \kq_{\rm CP}$ no temperature PT is expected by the TPCA.
However, the behavior of $q(t)$ is characteristic of a quadrupolar
phase at low temperatures  and of a paramagnetic phase at high
temperatures.

Figure~\ref{fig1} also contains the line corresponding to the
maxima of the static magnetic susceptibility $\chi(t)$ and the
line of the inflection points of the $q(t)$ curve. The line of the
$\chi(t)$ maxima converges with the line of the quadrupolar
phase~\ra paramagnetic phase transition at the left side of the
critical point at $\kq=\kq_{\chi}$. At $\kq_{\chi}<\kq<\kq_{\rm
CP}$ $\chi(t)$ is not a decreasing function in the paramagnetic
phase, but has a maximum. The line of the $q(t)$ inflection points
converges with the line of the quadrupolar phase~\ra paramagnetic
phase transitions in the critical point at $\kq=\kq_{\rm CP}$.

The phase diagram obtained within mean field approximation is
qualitatively different \cite{Chen,UFG}. At $\kq < 1.5$  and
$1.5~<~\kq~<~3.0$ the MFA (similarly to TPCA) predict the temperature
ferromagnetic~\ra paramagnetic phase transitions of the second
and first order, respectively ($\kq =1.5$ and $\kq =3.0$
are the coordinates of the tricritical and triple points).
However, within the MFA the $(\kq,t)$ phase diagram of
model (\ref{f1}) in zero magnetic field case contains
no critical point: the first order PT quadrupolar~\ra paramagnetic
phase is predicted by MFA at any $\kq>3.0$.

It should be mentioned that such a critical point may appear
within the MFA in a more general Blume-Emery-Griffiths model \cite{ri1}.

In the non-zero magnetic field case, similarly to the zero field
case, we shall distinguish the three following phases:
ferromagnetic, ``paramagnetic'', and ``quadrupolar''. The
``paramagnetic'' and ``quadrupolar'' phases differ from the
paramagnetic and quadrupolar phases by non-zero field-induced
magnetization only. The $m(t)$ is an increasing function in the
``quadrupolar'' phase and a decreasing convex downward function in
the ``paramagnetic'' phase. In non-zero magnetic field the
temperature PT can be of the first order only.

\begin{figure}
\begin{center}
{\includegraphics*{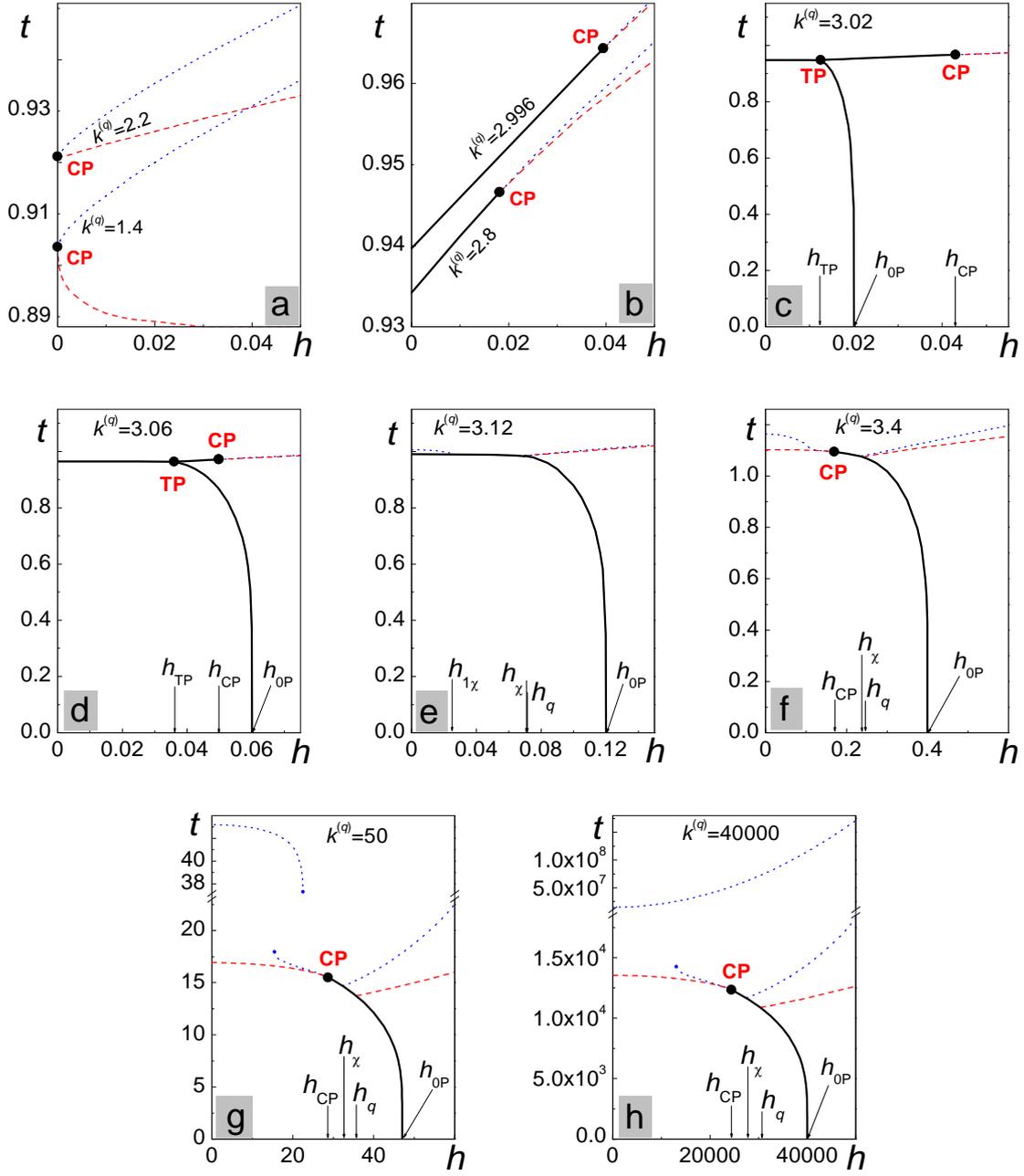}}\\ [1ex]
\end{center}
\caption{ The $h$ vs $t$ phase diagrams at different values of
quadrupolar interaction: a~-- $\kq=$1.4, 2.2; b~-- 2.8, 2.996; c~--
$\kq=3.02$; d~-- $\kq=3.06$; e~-- $\kq=3.12$; f~-- $\kq=3.4$; g~--
$\kq=50$; h~-- $\kq=40000$.
Thick solid line indicates the first order temperature
PT. Thin dashed and dotted lines correspond to the inflections in
the temperature dependences of quadrupolar moment and to the
maxima in the temperature dependences of static magnetic
susceptibility, respectively. The special points are the triple
point (TP), critical point (CP), and
phase boundary point in the ground state (0P).
$h_{1 \chi}$, $h_{\chi}$ and $h_{q}$ are the coordinates of the
intersection points of temperature PT line and of
the lines, which correspond to maxima in $\chi(t)$ and to
inflections in $q(t)$. } \label{fig2}
\end{figure}

Figure~\ref{fig2} shows the obtained within the  TPCA  $(h,t)$
phase diagrams at different values of the quadrupolar interaction.
The diagrams also contain the lines corresponding to the maxima of
the static magnetic susceptibility and to the inflection points of
the quadrupolar moment temperature curves.
Figure~\ref{fig2} illustrates the major aspects of the changes in
the topologies of the  $(h,t)$ phase diagrams with changing $\kq$.

At those values of the quadrupolar interaction, when at $h=0$ the
system undergoes the second order ferromagnetic~\ra paramagnetic
phase transition on increasing temperature ($\kq < \kq_{\rm
TCP}$), the magnetic field leads to ``smearing out'' (disappearance)
of the temperature phase transition. Instead of  vanishing
magnetization and instead of the cusp in the quadrupolar moment
temperature curve (as at the second order ferromagnetic~\ra
paramagnetic phase transition), magnetic field induces inflection
points in the temperature curves of magnetization and quadrupolar
moment, whereas the $\chi(t)$ does not diverge but has a finite
maximum. Topology of the $(h,t)$ phase diagrams at $\kq \in [0,
\kq_{\rm TCP}[$ are illustrated in fig.~\ref{fig2}a.
These diagrams contain critical points. The points
of the second order PT ferromagnetic~\ra
paramagnetic phase at $h=0$ are the critical points in the $(h,t)$
diagrams ($h_{\rm CP}=0$, $t_{\rm CP}=t_c$, where $t_c$ is the
transition temperature).

The typical $(h,t)$ phase diagrams for those values of quadrupolar
interaction when at $h=0$ the system undergoes the first order
ferromagnetic~\ra paramagnetic phase transition on increasing
temperature ($\kq_{\rm TCP}<\kq < \kq_{\rm TP}$) are given in
fig.~\ref{fig2}b. At $h<h_{\rm CP}$
the system undergoes the first order transition ferromagnetic~\ra
``paramagnetic'' phase. The larger $h$, the lower are jumps of the
thermodynamic characteristics at this transition. At $h=h_{\rm
CP}$ the jumps vanish, and at $h>h_{\rm CP}$ no temperature PT
takes place.

However, at those values of the quadrupolar interaction when the
system at $h=0$ undergoes the first order transition
quadrupolar~\ra paramagnetic phase
($\kq_{\rm TP}<\kq < \kq_{\rm CP}$), the
changes taking place with the magnetic field are not as clear as
in the two described above cases. The  $(h,t)$ phase diagrams can
be of three different topologies (see figs. \ref{fig2}c --
\ref{fig2}e) in the three parts of the $\kq\in]\kq_{\rm TP},
\kq_{\rm CP}[$ interval: $\kq\in] \kq_{\rm TP}, 3.048[$, $\kq\in]3.048,
3.099[$, $\kq\in]3.099, \kq_{\rm CP}[$.

Topology of the $(h,t)$ phase diagram at $\kq\in] \kq_{\rm TP}, 3.048[$
is shown in the fig.~\ref{fig2}c. The diagrams contain triple points
at $h=h_{\rm TP}$, critical points at $h=h_{\rm CP}$, and the
ground state phase boundary points (0P) at $h=h_{\rm
0P}=\kq-3$, where $h_{\rm 0P}<h_{\rm CP}$. At low fields, the
system undergoes the temperature transition ``quadrupolar''~\ra
``paramagnetic'' phase. Increasing field splits this PT at
the triple point into the cascade of the transitions
``quadrupolar''~\ra ferromagnetic~\ra ``paramagnetic'' phase. Further
increasing of field decreases the temperature of the
``quadrupolar''~\ra ferromagnetic phase transition down to its
vanishing  at zero temperature at $h=h_{\rm 0P}$ and decreases
the jumps of the thermodynamic characteristics at the temperature
transition ferromagnetic~\ra ``paramagnetic'' phase down to their
vanishing and ``smearing out'' of the  transition.

Topology of the $(h,t)$ phase diagram at $\kq\in]3.048, 3.099[$
shown in fig.~\ref{fig2}d differs from the described above ones by
the fact that $h_{\rm CP}<h_{\rm 0P}$. The system undergoes the
temperature PT ``quadrupolar''~\ra ``paramagnetic'' phase at
$h\in]0,h_{\rm TP}[$, a cascade of the transitions
``quadrupolar''~\ra ferromagnetic~\ra ``paramagnetic'' phase at
$h\in]h_{\rm TP},h_{\rm CP}[$, and the transition
``quadrupolar''~\ra ferromagnetic phase at $h\in]h_{\rm CP},h_{\rm
0P}[$.

It should be also noted that the lines corresponding to the maxima
of $\chi(t)$ and to the inflection points of  $q(t)$  in all the
described above cases (see figs.~\ref{fig2}a -- \ref{fig2}d)
converge into the critical point.

At $\kq=3.099$ the triple and critical points at the $(h,t)$ phase
diagram coalesce and disappear along with the line of the
ferromagnetic~\ra ``paramagnetic'' phase transition; at $\kq
> 3.099$ there is no longer a boundary between the ferromagnetic and
``paramagnetic''
phases. At $\kq\in]3.099, \kq_{\rm CP}[$  the topology of the
phase diagrams is the same as of the diagram given in
figure~\ref{fig2}e. Here the lines corresponding to the maxima of
$\chi(t)$ converge with the line of the phase transitions at
$h=h_{1 \chi}$ and $h=h_{\chi}$, whereas the line of the
inflection points of $q(t)$ converges with the PT
line at $h=h_{q}$; here $h_{\chi} \ne h_{q}$. At $h\in[0,h_{\rm
0P}[$ the system undergoes a temperature PT; at
small fields the temperature dependences of thermodynamic
characteristics in the vicinity of $t_c$ are the same as at the
``quadrupolar''~\ra ``paramagnetic'' phase transition, whereas at
fields close to $h_{\rm 0P}$~ they are as at the
``quadrupolar''~\ra ferromagnetic phase transition. At fields close
to  $h_{q}$ in the high-temperature phase near $t_c$ the temperature
dependences of some thermodynamic characteristics are as in the
``paramagnetic'' phase,
whereas the dependences of the others are as in the ferromagnetic
phase.

At $\kq > \kq_{\rm CP}$, when at zero magnetic field there is no
temperature PT in the system, the topology of the
$(h,t)$ phase diagrams are as shown in figs.~\ref{fig2}f --
\ref{fig2}h. Here the critical points  at $h=h_{\rm CP}$ and the
ground state phase boundary points at  $h=h_{\rm 0P}$ are
present. Increasing $\kq$ does not qualitatively change the
phase diagram. However, the curve corresponding to the maxima
of $\chi(t)$ at $h<h_{\rm CP}$ splits into two branches
(see figs.~\ref{fig2}f -- \ref{fig2}h).

The critical points at $\kq > \kq_{\rm CP}$
are of a different type than those at
the phase diagrams shown in fig.~\ref{fig2}a
and in figs.~\ref{fig2}b -- \ref{fig2}d. In
these critical points the temperature phase transition arises on
increasing field, not disappears. At rather small fields from the
interval $]h_{\rm CP}, h_{\rm 0P}[$, the behavior of the
thermodynamic characteristics near $t_c$ is as at the
``quadrupolar''~\ra ``paramagnetic'' phase transition; at sufficiently
high fields the behavior is as at the ``quadrupolar''~\ra
ferromagnetic phase transition. It should be also noted that the
lines corresponding to the maxima of $\chi(t)$ and inflection
points of $q(t)$ at low fields converge with the line of the phase
transition at the critical point (at $h=h_{\rm CP}$) and at high
fields converge with the line of the PT at several
points ($h_{\chi}<h_{q}$).

\section{Conclusions}

Within the two-particle cluster approximation, we study the spin-1
Ising model with bilinear and quadrupolar interactions
in a magnetic field for the simple cubic lattice. It is shown that
at those values of quadrupolar interactions  when at zero field
the system undergoes a first or second order temperature phase
transition from the ferromagnetic to paramagnetic phase, the phase
diagrams in the (temperature, magnetic field) plane contain a
critical point. At those values of the quadrupolar interactions
when at zero field the system undergoes a first  order temperature
phase transition from the quadrupolar to paramagnetic phase, the
phase diagrams in the (temperature, magnetic field) plane may
contain a triple point, a critical point, and a phase boundary
point in the ground state or a  phase boundary point in the ground
state  only. At those values of the quadrupolar interactions when
at zero field there is no temperature phase transition, the phase
diagrams in the (temperature, magnetic field) plane contain a
critical point and a phase boundary point in the ground state.

It is also shown that the studied model has three types of the
critical points in the (temperature, magnetic field) phase
diagrams. At the critical points of the first type, there is a
second order phase transition from the ferromagnetic to
paramagnetic phase at zero field; application of field smears out
this transition. At the critical points of the second type, the
first order temperature phase transition  from the ferromagnetic
to ``paramagnetic'' phase disappears on increasing field. At the
critical points of the third type, a transition arises
on increasing field; in this
case the behavior of the thermodynamic characteristics near the
transition temperature is as at the first order phase transition
from  the ``quadrupolar'' to ``paramagnetic'' phase.

It is established that if the phase diagram in the (temperature,
magnetic field) plane contains a critical point, then at this
point the line
 of phase transition and the lines corresponding to the
inflections and maxima of the temperature curves of the
quadrupolar moment and static magnetic susceptibility,
respectively, do converge.

\clearpage
\onecolumn

\end{document}